\documentclass[12pt]{article}
\begin{document}
\input epsf
\def\be{\begin{equation}}
\def\bea{\begin{eqnarray}}
\def\ee{\end{equation}}
\def\eea{\end{eqnarray}}
\def\d{\partial}
\def\la{\lambda}
\def\eps{\epsilon}
\newcommand{\dm}{\begin{displaymath}}
\newcommand{\edm}{\end{displaymath}}
\renewcommand{\b}{\tilde{B}}
\newcommand{\gm}{\Gamma}
\newcommand{\ac}[2]{\ensuremath{\{ #1, #2 \}}}
\renewcommand{\ell}{l}
\def\bb{$\bullet$}

\def\q{\quad}

\def\bn{B_\circ}

\let\a=\alpha \let\b=\beta \let\g=\gamma \let\d=\delta \let\e=\epsilon
\let\z=\zeta \let\c=\chi \let\th=\theta  \let\k=\kappa
\let\l=\lambda \let\m=\mu \let\n=\nu \let\x=\xi \let\r=\rho
\let\s=\sigma \let\t=\tau
\let\vp=\varphi \let\vep=\varepsilon
\let\w=\omega      \let\G=\Gamma \let\D=\Delta \let\Th=\Theta
             \let\P=\Pi \let\S=\Sigma

\def\nn{\nonumber}
\let\bm=\bibitem

\let\pa=\partial

\begin{flushright}
OHSTPY-HEP-T-03-004\\
hep-th/0304007
\end{flushright}
\vspace{20mm}
\begin{center}
{\LARGE  Tachyon condensation and `bounce' in the D1-D5 system}
\\
\vspace{20mm}
{\bf  Oleg Lunin${}^1$,   Samir D. Mathur${}^2$, I.Y.  Park${}^2$ and
Ashish Saxena${}^2$
\\}
\vspace{4mm}
${}^1$School of Natural Sciences, Institute for Advanced Study,\\
Princeton, NJ 08540, USA\\
\vspace{4mm}
${}^2$Department of Physics,\\ The Ohio State University,\\ Columbus,
OH 43210, USA\\
\vspace{4mm}
\end{center}
\vspace{10mm}
\begin{abstract}
We construct supergravity solutions dual to microstates of the
D1-D3-D5 system with nonzero B field
moduli. Just like the  D1-D5 solutions in  hep-th/0109154 these
solutions are generically nonsingular
everywhere, with the `throat' closing smoothly near r=0.  We write  expressions
relating the asymptotic supergravity fields to the integral brane
charges. We study the infall of a D1
brane  down the throat of the geometries. This test brane `bounces'
off the smooth end for generic initial
conditions.  The details of the bounce depend on both the choice of
D1-D3-D5 microstate and the direction
of approach of the infalling D1 brane. In the dual field theory
description we see that the tachyon mode
starts to condense, but the tachyon bounces back up the potential
hill without reaching the deepest point
of the potential.

\end{abstract}
\newpage

\section{Introduction}
\renewcommand{\theequation}{1.\arabic{equation}}
\setcounter{equation}{0}

\subsection{Motivation}

             The idea of $AdS/CFT$ duality \cite{maldacena} suggests
that we make a radical change in our notion of matter and spacetime.
Consider a collection
of branes placed in asymptotically flat space. The spacetime near
the branes will be
deformed; let us choose the branes such that the metric is
$AdS_m\times S^n$ in the
near horizon limit.  In Einstein's picture of gravity we have matter
(the branes) near
$r=0$, and a consequent metrical deformation near $r=0$. But the $AdS/CFT$
correspondence says that the branes are {\it dual} to the near
horizon geometry. This
suggests that if we talk about the branes as well as the near horizon
$AdS$ geometry
then we are `double-counting'. In particular if we follow the curved
metric down to the
vicinity of
$r=0$ then we should {\it not} find the branes there. Is this
conclusion correct, and can
we explicitly observe this absence of branes in the full string
theory solution?

This question was addressed in \cite{lm4} where we considered the
D1-D5 system \cite{stromingervafa,callanmalda}.
             Consider type IIB string theory in flat  space, and compactify 5
spatial directions on  $T^4\times S^1$. We wrap $n_1$ D1 branes on
the $S^1$ and
$n_5$ D5 branes on $T^4\times S^1$. The near horizon geometry is
$AdS_3\times S^3\times T^4$.

The D1-D5 brane system is in the Ramond (R) sector, and it has $\sim
e^{2\sqrt{2}\pi\sqrt{n_1n_5}}$ degenerate ground states. It was found
that these
different `matter ground states' corresponded to different dual
geometries -- each
geometry was flat at infinity and had a throat that was locally
approximately $AdS_3\times S^3\times T^4$, but the throat ended in a
shape that was
different for different microstates. Further, in the CFT we can
compute a time period
$\Delta t_{CFT}$ for a pair of excitations on the CFT state to travel
once around the
`effective string' describing the D1-D5 bound state. This time was
found to exactly equal
the time $\Delta t_{SUGRA}$ taken for a supergravity quantum to
travel down the throat
and reflect back up from the end. It was crucial that the quantum in the latter
computation did not encounter any `matter' at the end of the throat
where it could be
trapped for a further length of time. Thus all the properties of the
`matter state' were
encoded in the depth and shape of the `throat' in the  geometry dual
to the state.

Different geometries in the above system were characterized by
different shapes of a
`singular curve' at the end of the throat. One might think that this
singular curve was
somehow the location of the D1-D5 branes that made up the geometry,
but it was shown
\cite{lm4} that the singularity here was `mild' in the sense that all
incoming waves
reflected off the singularity instead of entering it and getting
`trapped'. It was observed in
\cite{lmm} that this `mild singularity' was in fact just a coordinate
singularity in the
generic solution, similar to the singularity at the core of a
Kaluza-Klein monopole
\cite{grossperry}. Thus the geometries dual to the different D1-D5
states are in fact
generically completely nonsingular, and singularities that arise in
degenerate cases are
just those that occur when two Kaluza-Klein monopoles approach each
other. This makes
the generic case similar to a special system studied in \cite{bal,
mm} where it was found
that the maximally rotating D1-D5 system was described by a
nonsingular geometry. We  rename the `singular curve' of \cite{lm4} the
`central curve' of the geometry, since points on this curve are the centers of
        Kaluza-Klein monopoles which expand out in the  directions
transverse to the curve.

The above results have yielded significant progress in our
understanding of how the
$AdS/CFT$ correspondence works. We would now like to further explore
the nature of this
correspondence, by investigating other phenomena in the field theory and
observing their dual
behavior in the string solution.  The D1-D5 system has a finite
dimensional moduli space.
The gravity solutions constructed in \cite{lm4} were obtained for a
special subspace of
this moduli space, where all gauge potentials on the $T^4$ were set
to zero. For this special
subspace we have the property that the D1-D5 system is `threshold
bound'; thus we can
separate away some D1 and D5 branes away from the D1-D5 bound state
at no cost in
energy.

At generic points in moduli space, however, there is a binding energy
that prevents such
a separation. If we start with a set of branes that are {\it not}
bound to the remainder,
then we expect to get a tachyonic open string mode  on the system,
and condensation of
this tachyon would lower the energy and yield the actual bound state.
To study the
gravity dual of this phenomenon, we need to construct supergravity
solutions describing the D1-D5 bound state at values of the
moduli where the binding energy is nonzero. We construct a class of
such solutions in this
paper, and then use then to study the evolution of the tachyon in the
dual field theory.

\subsection{The solutions}

We construct geometries with the following properties:

\bigskip

(a)\q Spacetime is compactified on $T^4\times S^1$. Let the $T^4$ be
rectangular, and parametrized by coordinates $z_1, z_2, z_3, z_4$.
\bigskip

(b)\q We have $n_1$ D1 branes wrapped on $S^1$, $n_5$ D5 branes wrapped on
$T^4\times S^1$.
\bigskip

(c)\q We have $n_{12}$ D3 branes wrapped on $S^1$ and the directions
$z_1, z_2$ of $T^4$,
and $n_{34}$ D3 branes wrapped on $S^1$ and the directions $z_3,
z_4$ of $T^4$.
\bigskip

(d)\q We have a nonzero value at infinity for $b_{12}\equiv B_{z_1z_2}$,
$b_{34}\equiv B_{z_3,z_4}$, where
$B_{\mu\nu}$ is the NS-NS 2-form gauge field.
\bigskip

(e)\quad The generic geometry of the family is smooth everywhere, including the
interior near
$r=0$.
\bigskip

Solutions of having the properties (a)-(d) were constructed in
\cite{costaPerry,malrusso,wadia}.  But the harmonic
functions involved in the solutions had $1/r^2$ singularities.  We
follow the procedure of
\cite{HashNc,malrusso} to construct geometries with the additional 
property
(e), and this is done by
starting with the solutions of \cite{lm4} which, as mentioned above,
are  generically smooth in the
interior.  It was argued in \cite{lm4} that true
bound states of the
D1-D5 system have throats that end  due to the nonzero
size of the D1-D5
bound state, and the naive solution written using harmonic functions
with ${1\over r^2}$
singularity did not represent any configuration of the actual D1-D5
system.

After finding these solutions, we compute their mass and charges from
their asymptotic
behavior. We derive general expressions to count the numbers of
different kinds of branes from the asymptotic values of the gravity fields.  We
thus find the mass $M_{sugra}(n_1, n_5,  n_{12}, n_{34},
b_{12},b_{34})$ of the supergravity solution
       as a function of the numbers of branes present in the bound state.

We next look at the field theory description of the  bound state of
branes for the same brane charges
and moduli, and
write down the mass  $M_{CFT}(n_1, n_5, n_{12}, n_{34},
b_{12},b_{34})$ expected for the bound state if we
assume that the state is BPS.  We then perform a computation along the lines of
\cite{taylormethod} to show that there is indeed a supersymmetric
bound state with these
charges, at least at the level of the classical brane action. We find
$M_{CFT}=M_{sugra}$, as expected.

\subsection{Tachyon condensation and bounce}

We next turn to the computation that we wish to pursue-- the infall
of a D1 brane towards the D1-D3-D5 bound
state.  When there were no D3 branes and $B$ was zero then the D1
brane  felt no force of attraction towards the
D1-D5 bound state, since the D1-D5 system was threshold bound. With
$B\ne 0$ we will find an attractive force
on the D1 brane, and this `test brane' starts to fall down the throat
of the supergravity solution. In the dual
`brane' description we expect that this infall is described by a
process of tachyon condensation \cite{senreview}\footnote{For earlier 
work on tachyon condensation see for example \cite{halpern}.}.

A similar computation (for small values of $B$) was performed in
\cite{wadia}, but there the throat was an infinite
one, and the D1 brane proceeded down this throat without returning.
But we have argued that the correct duals
of  D1-D3-D5 bound states have throats that are closed at the end, and
this leads us to the phenomenon that we
wish to study. What happens to the infalling D1 brane when it reaches
the end of the throat?

We find that generically the D1 brane `bounces off' the end of the
throat.  The details of the bounce and
subsequent evolution depend on the choice of D1-D3-D5 bound state
(which determined the shape of the end of
the throat) as well as the direction of infall of the D1 brane. For a
special class of initial conditions the D1 brane
settles down, as $t\rightarrow\infty$, to a point on the `central
curve' of the geometry mentioned above, and
becomes in the process an `ordinary graviton' (as opposed to a giant
graviton) traveling at the speed of light.

\section{Constructing the supergravity solutions}
\renewcommand{\theequation}{2.\arabic{equation}}
\setcounter{equation}{0}

We start with the D1-D5 solutions constructed in \cite{lm4}.  Let us
briefly recall how
these solutions were found. We can map the D1-D5 system, by a sequence of S,T
dualities, to the FP system, where we have a fundamental string (F)
wrapped $n_5$ times
on the $S^1$ and $n_1$ units of momentum charge (P) also
along the $S^1$. The
bound state of these charges has the F string in the form of a single
multiply wound
string, and all the momentum P is carried by traveling waves on the
string. The
supergravity solution for such a multiply wound string can be
constructed \cite{lm3},
\cite{lm4}, by superposing the harmonic functions  arising from
different strands \cite{dabholkar,
callan}. In  carrying the P charge the F string is forced
to bend in the
transverse directions, and the bound state thus acquired a nonzero
size. The fact that
many profiles of the F string carry the same P leads (after
quantization) to the large
degeneracy
$\sim
e^{2\sqrt{2}\pi\sqrt{n_1n_5}}$ of ground states.  The classical
solutions for the FP bound
states are parametrized by the transverse  displacement profile $\vec
F(v)$ of the F
string. Undoing the S,T dualities we obtain the D1-D5 geometries,
still parametrized by
this profile function:
\bea\label{fiveMetr}
ds^2&=&\sqrt{\frac{H}{1+K}}\left[-(dt-A_idx^i)^2+(dy+B_idx^i)^2\right]
+\sqrt{\frac{1+K}{H}}
d{\vec x}d{\vec x}\nonumber\\
&+&\sqrt{H(1+K)}\left[(dz_1^2+dz_2^2)+(dz_3^2+dz_4^2)\right]\\
e^{2\Phi}&=&H(1+K)\quad \\
C^{(2)}_{ty}&=&-\frac{K}{1+K}, \quad
C^{(2)}_{iy}=-\frac{A_i}{1+K},\quad
C^{(2)}_{ti}=\frac{B_i}{1+K},\nonumber\\
C^{(2)}_{ij}&=&C_{ij}+\frac{A_iB_j-B_iA_j}{1+K}
\label{five}
\eea
where $H^{-1}$, $K$ and $A_i$ are given in terms of the string profile:
\be\label{profile}
H^{-1}=1+{Q_5\over L}\int_0^L{dv\over|{\bf x}-{{\bf F}}|^2},\quad
K={ Q_5\over L}\int_0^L{|\dot F|^2dv\over
|{\bf x}-{{\bf F}}|^2},\quad A_i=-{Q_5\over L}\int_0^L{{\dot
F}_i dv\over|{\bf x}-{{\bf F}}|^2}
\ee
and the forms $B$ and $C$ are defined by
\be
dC=-^{*_4}~dH^{-1},\qquad dB=-^{*_4}~dA
\ee
Here the dual $*_4$ is taken in the 4-dimensional Euclidean space $x_1,
x_2, x_3, x_4$ equipped with the flat metric $dx_idx_i$.

        Now we wish to find
solutions where the NS--NS two--form field has nonvanishing components
$B_{z_1z_2}, B_{z_3z_4}$ at infinity.  We can add in the above solution
(\ref{five})   constant values for these gauge fields
\be
B_{z_1z_2}=b_{12}, ~~~ B_{z_3z_4}=b_{34}
\label{one}
\ee
This makes $B\ne 0$ at infinity, but also adds D3 branes to the
system in such a way that the overall
mass is unchanged. To obtain another independent combination of $B$
and D3 charge we follow the
procedure of \cite{HashNc,malrusso}\footnote{For earlier applications of 
the
similar methods to constructing supergravity solutions see \cite{russo}.}
: We do a T-duality along $z_1$,
perform a rotation in the $z_1-z_2$
plane, and then T-dualize again in  $z'_1$ -- this gives nonvanishing
D3 charge and a
non--constant $B_{z_1z_2}$ (which vanishes at infinity).  A similar
procedure is performed using the
coordinates $z_3, z_4$. The steps of this calculation are given in
Appendix A, and we summarize the result
here:
\bea\label{NCSolution}
ds^2&=&\sqrt{\frac{H}{1+K}}\left[-(dt-A_idx^i)^2+(dy+B_idx^i)^2\right]
+\sqrt{\frac{1+K}{H}}
d{\vec x}d{\vec x}\nonumber\\
&+&\sqrt{H(1+K)}\left[h_1^{-1}(dz_1^2+dz_2^2)+h_2^{-1}(dz_3^2+dz_4^2)\right]\\
e^{2\Phi}&=&\frac{1}{h_1h_2}H(1+K),\quad
B_{z_1z_2}=\frac{\sin\theta_1\cos\theta_1}{h_1}(1-(1+K)H)+b_{12}
,\\
&&B_{z_3z_4}=\frac{\sin\theta_2\cos\theta_2}{h_2}(1-(1+K)H)+b_{34},\\
C^{(2)}_{ty}&=&-\frac{K\cos\theta_1\cos\theta_2}{1+K}\quad
C^{(2)}_{iy}=-\frac{A_i\cos\theta_1\cos\theta_2}{1+K}, \quad
C^{(2)}_{ti}=\frac{B_i\cos\theta_1\cos\theta_2}{1+K},\nonumber\\
C^{(2)}_{ij}&=&\cos\theta_1\cos\theta_2\left(C_{ij}+\frac{A_iB_j-B_iA_
j}{1+K}\right),\\
C^{(4)}_{ty34}&=&-\frac{KH}{h_2}\cos\theta_1\sin\theta_2,\quad
C^{(4)}_{ty12}=-\frac{KH}{h_1}\cos\theta_2\sin\theta_1,\nonumber\\
C^{(4)}_{iy34}&=&-\frac{HA_i}{h_2}\cos\theta_1\sin\theta_2,\quad
C^{(4)}_{iy12}=-\frac{HA_i}{h_1}\cos\theta_2\sin\theta_1,\nonumber\\
C^{(4)}_{it34}&=&-\frac{HB_i}{h_2}\cos\theta_1\sin\theta_2,\quad
C^{(4)}_{it12}=-\frac{HB_i}{h_1}\cos\theta_2\sin\theta_1,\nonumber\\
C^{(4)}_{ij34}&=&\frac{H(1+K)}{h_2}\cos\theta_1\sin\theta_2\left(C_{ij
}+\frac{A_iB_j-B_iA_j}{1+K}\right),
\nonumber\\
C^{(4)}_{ij12}&=&\frac{H(1+K)}{h_1}\cos\theta_2\sin\theta_1\left(C_{ij
}+\frac{A_iB_j-B_iA_j}{1+K}\right),
\nonumber\\
C^{(6)}_{ty1234}&=&-\frac{H^2K(1+K)\sin\theta_1\sin\theta_2}{h_1h_2},\quad
C^{(6)}_{iy1234}=-\frac{H^2(1+K)A_i\sin\theta_1\sin\theta_2}{h_1h_2},
\nonumber\\
C^{(6)}_{it1234}&=&-\frac{H^2(1+K)B_i\sin\theta_1\sin\theta_2}{h_1h_2}
,\nonumber\\
C^{(6)}_{ij1234}&=&\frac{H^2(1+K)^2\sin\theta_1\sin\theta_2}{h_1h_2}
\left(C_{ij}+\frac{A_iB_j-B_iA_j}{1+K}\right),\nonumber
\label{six}
\eea
where
\be
h_i=\cos^2\theta_i+\sin^2\theta_i(1+K)H, ~ ~~i=1,2
\label{seven}
\ee
and $b_{12}$, $b_{34}$ are the values of $B$ at $r\rightarrow\infty$.

\section{Mass and charges of the solution}
\renewcommand{\theequation}{3.\arabic{equation}}
\setcounter{equation}{0}

In this section we study the asymptotic behavior of the solutions
(\ref{NCSolution}) and derive their mass and
their charges. The charges are to be expressed as   integers that
give the numbers of D1,
D3 and D5 branes in the configuration. The extraction of these
integers is complicated by the fact
that the field $B$ is nonzero at infinity; a $p+2$-form field
strength contributes not only to the
count of $p$-branes but also to branes of other dimensionalities when
$B\ne 0$. Thus we begin with
a derivation of the relevant field-charge relations for the theory,
and then compute the charges for
our solution.

\subsection{Field equations}

Let us begin with the action for type IIB supergravity in the absence of
sources. We use the notation of
\cite{johnson}:
\bea\label{NewAction}
S_{IIB}&=&\frac{1}{(2\pi)^7{\alpha'}^4}\int \left\{d^{10}x\sqrt{-G}
e^{-2\Phi}\left[R+4(\nabla\phi)^2-\frac{1}{12}\left(H^{(3)}\right)^2\right]
\right.\nonumber\\
&-&\left.\frac{1}{2}~^*F^{(3)}\wedge F^{(3)}-
\frac{1}{2}~^*dC^{(0)}\wedge dC^{(0)}-
\frac{1}{4}~^*F^{(5)}\wedge F^{(5)}\right\}\nonumber\\
&+&\frac{1}{2(2\pi)^7{\alpha'}^4}\int \left(C^{(4)}+\frac{1}{2}B^{(2)}
\wedge
C^{(2)}
\right)
\wedge G^{(3)}\wedge H^{(3)}.
\eea
Here
\be
G^{(3)}=dC^{(2)},\quad
F^{(3)}=dC^{(2)}+C^{(0)}H^{(3)},\quad F^{(5)}=dC^{(4)}+
H^{(3)}\wedge C^{(2)}
\ee

In this convention the four form $C^{(4)}$ is invariant under the gauge
transformation of $ B^{(2)}$, while under the gauge transformation of the two
form ${\delta C^{(2)}}$ it transforms as
\be
{\delta C^{(4)}}=-B^{(2)}\wedge \delta C^{(2)}
\label{C2-TF}
\ee
For our solutions $C^{(0)}=0$, and we assume the vanishing of
$C^{(0)}$ in the equations
below.  We find the equations of motion following from the action
(\ref{NewAction}) by taking the variations with respect to $C^{(2)}$ and
$C^{(4)}$:
\bea
\label{EMotthree}
d^*F^{(5)}+ H^{(3)}\wedge F^{(3)}&=&0\\
\label{EMotone}
d^*F^{(3)}-H^{(3)}\wedge F^{(5)}&=&0
\eea
These equations should be supplemented by the Bianchi identity:
\be
dF^{(3)}=0
\ee
and the self duality condition $F^{(5)}=^*F^{(5)}$.

In the presence of sources we get a Chern-Simons coupling between
the RR gauge fields and the
currents
$j^{(p+1)}$ describing D-branes.
\bea
S_{CS}&=&\frac{g}{2\pi\alpha'}\int
\exp(B+2\pi\alpha' F)\wedge\sum_{p=2}^6 C^{(p)}\wedge \left[^*j^{(2)}
+\frac{1}{(2\pi)^2\alpha'}~^*j^{(4)}+\frac{1}{(2\pi)^4{\alpha'}^2}~^*j^{(6)}
\right]\nonumber
\eea
In this normalization when $F=0$ the number of branes is given by
integrating the
corresponding current over an appropriate cycle:
\be\label{charge}
n_k=\int ~^*j^{(k+1)}
\ee
(The $n_k$ are integers in the full quantum theory, while the value
of $B$ is a continuous variable
\cite{taylor}.)
The equations of motion are modified to
\bea\label{CurrentOne}
&&-\frac{1}{4\pi^2 g{\alpha'}}dF^{(3)}=^*j^{(6)}\nonumber\\
&&\frac{1}{4\pi^2 g{\alpha'}^2}\left[
d^*F^{(5)}+H^{(3)}\wedge F^{(3)}\right]=
(2\pi)^2~^*j^{(4)}+\frac{1}{\alpha'}(B+2\pi\alpha'F)\wedge^*j^{(6)}
\\
&&\frac{1}{4\pi^2g{\alpha'}^3}\left[d^* F^{(3)}-
H^{(3)}\wedge F^{(5)}\right]=
(2\pi)^4~^*j^{(2)}+\frac{(2\pi)^2}{\alpha'}(B+2\pi\alpha'F)\wedge
^*j^{(4)}\nonumber\\
&&\qquad\qquad\qquad+\frac{1}{2{\alpha'}^2}(B+2\pi\alpha'F)^2\wedge
^*j^{(6)}\nonumber
\eea

Assuming that there are no sources for the NS two form field (i.e.
$dH^{(3)}=0$), we rewrite (\ref{CurrentOne}) in a form which is more
convenient for the charge computation in supergravity:
\bea\label{CurrentTwo}
&&-\frac{1}{4\pi^2{\alpha'}g}dF^{(3)}=^*j^{(6)}\nonumber\\
&&\frac{1}{(4\pi^2{\alpha'})^2 g}d\left[
^*F^{(5)}+B^{(2)}\wedge F^{(3)}\right]=
^*j^{(4)}+\frac{F}{2\pi}\wedge^*j^{(6)}
\\
&&\frac{1}{(4\pi^2{\alpha'})^3 g}d\left[~^* F^{(3)}-
B^{(2)}\wedge F^{(5)}-
\frac{1}{2}B^{(2)}\wedge B^{(2)}\wedge F^{(3)}\right]=
^*j^{(2)}+\frac{F}{2\pi}\wedge ^*j^{(4)}\nonumber\\
&&\qquad\qquad\qquad
+\frac{1}{2}\left(\frac{F}{2\pi}\right)^2\wedge ^*j^{(6)}
\nonumber
\eea
In deriving  the last equation we have used the  relation
\bea
H^{(3)}\wedge F^{(5)}&=&d(B^{(2)}\wedge F^{(5)})-B^{(2)}\wedge(dF^{(5)}+
H^{(3)}\wedge F^{(3)})\nonumber\\
&+&\frac{1}{2}d(B^{(2)}\wedge B^{(2)}\wedge F^{(3)})-
\frac{1}{2}B^{(2)}\wedge B^{(2)}\wedge dF^{(3)}
\eea
as well first two equations in (\ref{CurrentOne}).

\subsection{Obtaining charges from the field strengths}

            From (\ref{CurrentTwo}) we read off the integer charges in
terms of the
field strengths
\bea
n_5&=&-\frac{1}{4\pi^2\alpha' g}\int_{S^3} F^{(3)}
\\
n_1&=&\frac{1}{(4\pi^2{\alpha'})^3g}\int_{S^3\times T^4}\left[
~^*F^{(3)}-B^{(2)}\wedge F^{(5)}-
\frac{1}{2}B^{(2)}\wedge B^{(2)}\wedge F^{(3)}\right]\\
n_{12}&=&\frac{1}{(4\pi^2{\alpha'})^2g}\int_{S^3\times T_3\times T_4}
(F^{(5)}+B^{(2)}\wedge F^{(3)})\\
n_{34}&=&\frac{1}{(4\pi^2{\alpha'})^2g}\int_{S^3\times T_1\times T_2}
(F^{(5)}+B^{(2)}\wedge F^{(3)})
\eea
Here $T_i, i=1,2,3,4$ are the four different cycles of $T^4$, and
$n_{12}$ for example gives the D3
branes wrapped on the cycles $T_1, T_2$.

Define
\bea
k_5&=&-\frac{1}{4\pi^2\alpha' g}\int_{S^3} F^{(3)}
\\
k_1&=&\frac{1}{(4\pi^2{\alpha'})^3g}\int_{S^3\times T^4}~^*F^{(3)}\\
k_{ij}&=&\eps_{ijkl}\frac{1}{(4\pi^2{\alpha'})^2g}
\int_{S^3\times T_k\times T_l}F^{(5)}
\eea

It is convenient to introduce
\be
b_{ij}\equiv \frac{1}{L_iL_j}\int _{T_i\times T_j}B
\ee
where $L_i$ is the length of the $i$--th direction on the torus.
We also define
\be
V_{ij}\equiv {L_i L_j\over (2\pi)^2}, ~~~V\equiv V_{12}V_{34}.
\ee

Then we get
\bea\label{BraneTransl}
n_5&=&k_5\nonumber\\
n_{12}&=&k_{12}-\frac{b_{34}V_{34}}{\alpha'}k_5,\qquad
n_{34}=k_{34}-\frac{b_{12}V_{12}}{\alpha'}k_5,
\\
n_1&=&k_1-\frac{b_{12}V_{12}}{\alpha'}k_{12}-
\frac{b_{34}V_{34}}{\alpha'}k_{34}+\frac{b_{12}b_{34}V}{(\alpha')^2}k_
5\nonumber
\eea

\subsection{Asymptotic behavior of the solution}

As $r\rightarrow\infty$ the solution (\ref{NCSolution}) has the following
behavior for the
fields
\bea
ds^2&=&-dt^2+dy^2+d{\vec x}d{\vec x}+dz_1^2+dz_2^2+dz_3^2+dz_4^2\\
e^{2\Phi}&=&1,\quad
B_{12}=\sin\theta_1\cos\theta_1\frac{Q_5-Q_1}{r^2},\quad
B_{34}=\sin\theta_2\cos\theta_2\frac{Q_5-Q_1}{r^2},\nonumber\\
C^{(2)}_{ty}&=&-\frac{Q_1}{r^2}\cos\theta_1\cos\theta_2\quad
C^{(2)}_{iy}=-A_i\cos\theta_1\cos\theta_2,
\quad
C^{(2)}_{ti}=B_i\cos\theta_1\cos\theta_2,\nonumber\\
C^{(2)}_{ij}&=&\cos\theta_1\cos\theta_2 C_{ij}\\
C^{(4)}_{ty34}&=&-\frac{Q_1}{r^2}\cos\theta_1\sin\theta_2,\quad
C^{(4)}_{ty12}=-\frac{Q_1}{r^2}\cos\theta_2\sin\theta_1,\nonumber\\
C^{(4)}_{iy34}&=&-A_i\cos\theta_1\sin\theta_2,\quad
C^{(4)}_{iy12}=-A_i\cos\theta_2\sin\theta_1,\nonumber\\
C^{(4)}_{it34}&=&-B_i\cos\theta_1\sin\theta_2,\quad
C^{(4)}_{it12}=-B_i\cos\theta_2\sin\theta_1,\nonumber\\
C^{(4)}_{ij34}&=&\cos\theta_1\sin\theta_2 C_{ij},\qquad
C^{(4)}_{ij12}=\cos\theta_2\sin\theta_1 C_{ij},
\nonumber\\
C^{(6)}_{ty1234}&=&-\frac{Q_1}{r^2}\sin\theta_1\sin\theta_2,\quad
C^{(6)}_{iy1234}=-A_i\sin\theta_1\sin\theta_2,\nonumber\\
C^{(6)}_{it1234}&=&-B_i\sin\theta_1\sin\theta_2,\qquad
C^{(6)}_{ij1234}=\sin\theta_1\sin\theta_2 C_{ij}\nonumber
\eea

Recall that the field strengths at infinity are to be constructed
from the potentials in such a way
that each field strength contains the information of both the
relevant electric charges as well as
their magnetic duals.
In computing the duals of forms we use the convention
\be
\eps_{tyijkl1234}=\eps_{ijkl}
\ee
where $i,j,k,l$ are the four noncompact directions $x_i$ and
$1,2,3,4$ are directions on the $T^4$.
For the D1 and D5 charges we need to compute $F^{(3)}$; since we are
working at infinity we can
drop nonlinear terms and we get
\be\label{ToProve}
F^{(3)}=dC^{(2)}-^*dC^{(6)}
\ee
(The relative sign of the two terms on the RHS is determined by
performing two T-dualities of the
5-form field strength, which is assumed to satisfy $F^{(5)}=*F^{(5)}$.)

At leading order we find
\bea
F^{(3)}&=&-Q_1\cos\theta_1\cos\theta_2 d\left(\frac{1}{r^2}\right)
\wedge dt\wedge dy+
\cos\theta_1\cos\theta_2 d{ C}\nonumber\\
&+&^*\left[-Q_1\sin\theta_1\sin\theta_2d\left(\frac{1}{r^2}\right)
\wedge dt\wedge dy\wedge dV
+\sin\theta_1\sin\theta_2 d{C}\wedge dV\right]
\eea
Note that at $r\rightarrow\infty$
\be
^*d{ C}=-Q_5d\left(\frac{1}{r^2}\right)\wedge dt\wedge dy\wedge dV
\ee
where $dV=dz_1\wedge dz_2\wedge dz_3 \wedge dz_4$. We  choose the orientation
of the $S^3$ at infinity to be given by the choice of sign in the
following relation
\be
\int_{S^3} d{C}=-4\pi^2 Q_5
\ee
    Then we arrive at the  result
\bea
F^{(3)}&=&-(Q_1\cos\theta_1\cos\theta_2+ Q_5\sin\theta_1\sin\theta_2)
             d\left(\frac{1}{r^2}\right)
\wedge dt\wedge dy\nonumber\\
&+&\frac{1}{Q_5}
(Q_5\cos\theta_1\cos\theta_2+ Q_1\sin\theta_1\sin\theta_2)
d{C},
\eea
\bea
^*F^{(3)}&=&-\frac{1}{Q_5}
(Q_1\cos\theta_1\cos\theta_2+ Q_5\sin\theta_1\sin\theta_2)
d{ C}\wedge dV \nonumber\\
&+&
(Q_5\cos\theta_1\cos\theta_2+ Q_1\sin\theta_1\sin\theta_2)
d\left(\frac{1}{r^2}\right)
\wedge dt\wedge dy\wedge dV
\eea
Substituting this in the expressions for the charges we find:
\bea
k_5&=&\frac{1}{g\alpha'}
(Q_5\cos\theta_1\cos\theta_2+ Q_1\sin\theta_1\sin\theta_2)
\\
k_1&=&\frac{V}{g(\alpha')^3}
(Q_1\cos\theta_1\cos\theta_2+Q_5\sin\theta_1\sin\theta_2)
\eea

Let us now evaluate the numbers of the three branes. First we have to find the
self--dual field strength. We define:
\be
F^{(5)}=dC^{(4)}+^*dC^{(4)}
\ee
At  leading order we find:
\bea
dC^{(4)}&=&-Q_1d\left(\frac{1}{r^2}\right)(\cos\theta_1\sin\theta_2 dV_{34}+
\cos\theta_2\sin\theta_1 dV_{12})\wedge dt\wedge dy\nonumber\\
&+&d{C}(\cos\theta_1\sin\theta_2 dV_{34}+\cos\theta_2\sin\theta_1
dV_{12})
\eea
\bea
F^{(5)}&=&-d\left(\frac{1}{r^2}\right)dV_{34}(Q_1\cos\theta_1\sin\theta_2-
             Q_5\cos\theta_2\sin\theta_1)\wedge dt\wedge dy\nonumber\\
&&-d\left(\frac{1}{r^2}\right)dV_{12}(
Q_1\cos\theta_2\sin\theta_1- Q_5\cos\theta_1\sin\theta_2)
\wedge dt\wedge dy\nonumber\\
&&+\frac{1}{Q_5}d{C}dV_{34}(Q_5\cos\theta_1\sin\theta_2-
Q_1\cos\theta_2\sin\theta_1)\nonumber\\
&&+\frac{1}{Q_5}d{C}dV_{12}(
Q_5\cos\theta_2\sin\theta_1- Q_1\cos\theta_1\sin\theta_2)
\eea
Integrating these expressions we get
\bea
k_{12}&=&-\frac{V_{34}}{g(\alpha')^2}(
Q_5\cos\theta_1\sin\theta_2- Q_1\cos\theta_2\sin\theta_1)\nonumber\\
k_{34}&=&-\frac{V_{12}}{g(\alpha')^2}
(Q_5\cos\theta_2\sin\theta_1- Q_1\cos\theta_1\sin\theta_2)
\label{qqtwone}
\eea

\subsection{Mass in terms of charges}
          From the asymptotic behavior of $g_{tt}$ we find the mass of
the solution
\be\label{massOne}
M=\frac{\pi}{4G_5}(Q_1+Q_5)=\frac{RV}{g^2{\alpha'}^4}(Q_1+Q_5)
\ee
We wish to relate this mass to the number of branes. We observe that
\be
V_{12}k_{12}- V_{34}k_{34}=
\frac{V_{12}V_{34}}{g{\alpha'}^2}
\sin(\theta_1-\theta_2)(Q_1+Q_5)
\ee
\be\label{kfive}
k_5+\frac{k_1(\alpha')^2}{V}=\frac{1}{g\alpha'}(Q_1+Q_5)
\cos(\theta_1-\theta_2)
\ee
Thus
\be
M^2=\left(\frac{RV}{g{\alpha'}^3}\right)^2
\left[(k_5+\frac{k_1(\alpha')^2}{V})^2+\frac{{\alpha'}^2}{V^2}
(V_{12}k_{12}- V_{34}k_{34})^2\right]
\label{qqone}
\ee

We express the quantities
$k_1, k_5, k_{12}, k_{34}$  in terms of the numbers of
different kinds of branes by the relations inverse to (\ref{BraneTransl}):
\bea
k_5&=&n_5\\
k_{12}&=&n_{12}+\frac{b_{34}V_{34}}{\alpha'}n_5,\qquad
k_{34}=n_{34}+\frac{b_{12}V_{12}}{\alpha'}n_5,
\\
k_1&=&n_1+\frac{b_{12}V_{12}}{\alpha'}n_{12}+
\frac{b_{34}V_{34}}{\alpha'}n_{34}+
\frac{b_{12}b_{34}V}{(\alpha')^2}n_5
\eea

\section{Mass of the D-brane state}
\renewcommand{\theequation}{4.\arabic{equation}}
\setcounter{equation}{0}

We have constructed above the gravity dual of a D1-D3-D5 bound state
with $B\ne 0$.  In this section we compute the mass
expected of such a state starting from D-brane physics. If we perform
a T-duality along the $S^1$ directions of $T^4\times S^1$,
then we get a $D0-D2-D4$ bound state; we study the latter since the
supercharges are easier to write for the IIA theory which can
in turn be written as a reduction of 11-dimensional M theory.

Consider a D0-D2 system. If the D0 brane is {\it not} bound to the D2
brane, the system is not supersymmetric; the
supersymmetries preserved by the D2 brane and the D0 brane are
different. If however we allow the  D0 brane to `dissolve' into
the D2 brane, then the mass  is lowered, and the bound state is a
supersymmetric (1/2 BPS ) configuration. The geometries we
have constructed are expected to be duals of the {\it bound} state of
the D1-D3-D5 branes, so we are interested in the masses of
such `dissolved' configurations.

If we {\it know}  that the bound state is supersymmetric, then we can
deduce the mass from the charges. Let the D0, D2, D4 charges be
described by $Z, Z_{ij}, Z_{ijkl}$. Then we write  \cite{towns,pioline}
\begin{equation}
\gm\epsilon\equiv Z \gm_{0s}+ \frac{1}{2!}Z^{ij}
\gm_{0ij}+\frac{1}{4!}Z^{ijkl}\gm_{0ijkls} = M \epsilon \label{qthree}
\end{equation}
Here the index $s$ represents the compact $11-$direction of M-theory,
while the other indices take values
along the compact torus. Requiring a solution $\epsilon\ne 0$ gives
$M$. This computation is standard, and for our
case of interest we reproduce the details in Appendix \ref{AppMass}.
T-dualizing to the D1-D3-D5 system we
get
\dm
M^{2}=
{R^2\over \alpha'}\frac{1}{g^2\alpha'}
\left(\left(\ell_{1}\ell_{2}n_{12}\mp
\ell_{3}\ell_
{4}n_{34}+ n_{5}v (b_{34}\mp b_{12})\right)^{2}+\right.
\edm
\be
\left.\left(n_{1}+ v(b_{12}b_{34}\pm 1)
n_{5}+\ell_{1}\ell_{2}b_{12}n_{12} +
\ell_{3}\ell_{4}b_{34}n_{34}\right)^{2}\right)
\label{qel}
\ee
where $\ell_{i} = \frac{L_{i}}{2\pi\sqrt{\alpha'}}$ are dimensionless
parameters expressing the lengths $L_i$ of the rectangular torus
$T^4$, $v= \ell_{1}\ell_{2}\ell_{3}\ell_{4}$ and the various $n$'s
are the number of respective branes. In the above expression
we choose the upper signs if $k_1k_5-k_{12}k_{34}>0$ and the
lower signs if
$k_1k_5-k_{12}k_{34}<0$.
We see that this expression for
the mass agrees
with  (\ref{qqone}).
(Note that
$V_{12}=\alpha' l_1l_2$,
$V_{34}=\alpha' l_3l_4$. Also note that in our supergravity
computation  we have chosen a definite sign for the charges
at the outset; this choice corresponds to the upper signs in  (\ref{qel}).)

As mentioned above, this would be the mass of the bound state {\it if}
we knew that the state was supersymmetric. But the question
of whether the bound state is supersymmetric or not is a dynamical
question in the theory, and this dynamical information is not
contained in the starting step (\ref{qthree}) of the above
computation. To determine whether the branes are actually expected
to
form a supersymmetric bound state we follow the approach used in
\cite{taylormethod}. In this approach we start with a
collection of $D4$ branes, represent the other (dissolved) branes in
terms of field strengths $F$ on the D4 branes, with the
assumption that $F$ can be taken as a diagonal $U(n_4)$ matrix.
Within this class of $F$ we check if there is a supersymmetric
configuration; if there is, then at least in a classical
approximation to brane physics we would establish that the bound
state is
supersymmetric.

We take a rectangular torus as above and let  $b_{12}, b_{34}$ be nonzero.
We have D0 and  D4 branes as well as D2 branes along the $12$ and
$34$ directions.
Since all quantities depend
only on $B+2\pi\alpha'F$ we can set $B=0$ and absorb its effect into
$F$ with no loss of
generality (the discreteness of $F$ is not visible in this classical
analysis). We label the D4 branes by an index $i$, with $i= 1\dots
n_4$.  The D4 branes carry field strengths $F_{12}^{(i)},
F_{34}^{(i)}$. Define the vectors
\be
\vec V_1=\{F_{12}^{(1)}, \dots , F_{12}^{(n_4)}\}
\label{qfour}
\ee
\be
\vec V_2=\{F_{34}^{(1)}, \dots , F_{34}^{(n_4)}\}
\label{qfourp}
\ee
\be
\vec V_0=\{ 1, \dots 1\}, ~~~V_0\cdot V_0=n_4
\label{qfourpp}
\ee
We have the constraints
\be
\sum_iF_{12}^{(i)}=\vec V_1\cdot\vec V_0=({2\pi\over L_1L_2})~ n_{34}
\label{qsix}
\ee
\be
\sum_iF_{34}^{(i)}=\vec V_2\cdot\vec V_0=({2\pi\over L_3L_4})~ n_{12}
\label{qseven}
\ee
\be
\sum_iF_{12}^{(i)}F_{34}^{(i)}=\vec V_1\cdot\vec V_2=({(2\pi)^2\over
L_1L_2L_3L_4} )~n_0
\label{qeight}
\ee
Lastly, we have the requirement that the different D4 branes be
supersymmetric with respect to each other. We can use either
             the Yang-Mills action to describe the branes or  the (more exact)
DBI action. In the Yang-Mills limit the
supersymmetry condition is \cite{douglas,mpt}
\be
F_{12}^{(1)}\pm F_{34}^{(1)}=F_{12}^{(2)}\pm F_{34}^{(2)}=\dots
=F_{12}^{(n_4)}\pm F_{34}^{(n_4)}
\label{qnine}
\ee
where the signs $\pm$ must be chosen to be all $+$ or all $-$. With
the DBI action we define
\be
f_{12}^{(i)}=\tan^{-1} {F_{12}^{(i)}\over 2\pi\alpha'},
~~f_{34}^{(i)}=\tan^{-1} {F_{34}^{(i)}\over 2\pi\alpha'}
\ee
and then the supersymmetry preservation condition is \cite{douglas,mpt}
\be
f_{12}^{(1)}\pm f_{34}^{(1)}=f_{12}^{(2)}\pm f_{34}^{(2)}=\dots
=f_{12}^{(n_4)}\pm f_{34}^{(n_4)}
\label{qthir}
\ee

Consider the Yang-Mills approximation. The constraint (\ref{qnine})
can be written as
\be
\vec V_1\pm\vec V_2= c ~\vec V_0
\label{qten}
\ee
If we can solve (\ref{qsix}), (\ref{qseven}), (\ref{qeight}),
(\ref{qten}) for real vectors $\vec V_i$ then we have a supersymmetric
configuration, and (\ref{qel}) would give the mass of the bound state.

The conditions (\ref{qsix}), (\ref{qseven}) are immediately solved by writing
\be
\vec V_1=c_1\vec V_0 + \vec V_1^{\perp}, ~~~c_1= {1\over
n_4}({2\pi\over L_1L_2})~ n_{34}
\ee
\be
\vec V_2=c_2\vec V_0 + \vec V_2^{\perp}, ~~~c_2= {1\over
n_4}({2\pi\over L_3L_4})~ n_{12}
\ee
where $\vec V^{\perp}_1\cdot \vec V_0=\vec V^{\perp}_2\cdot \vec
V_0=0$.  The condition (\ref{qten}) gives $\vec
V_1^\perp=\mp\vec V_2^\perp$,
and then (\ref{qeight}) gives
\be
|\vec V_1^\perp|^2=\pm~{1\over n_4} {(2\pi)^2\over
L_1L_2L_3L_4}~(n_{12}n_{34}-n_0n_4)
\label{qtw}
\ee
If $(n_{12}n_{34}-n_0n_4)\ge 0$  we take the $+$ sign in
(\ref{qnine}),  and then we can choose any $\vec V_1^\perp$ with real
entries and length given by (\ref{qtw}) to get a supersymmetric
configuration. If
$(n_{12}n_{34}-n_0n_4)\le 0$ then we can take the
$-$ sign and again get a supersymmetric configuration.\footnote{The
above computations assume $n_4>1$. We expect
supersymmetric  states also for $n_4=1$, but the state is not
described by a constant field $F$. For $n_4>1$ we have in general many
choices for the vector $\vec V_1^\perp$, and these choices reflect
the presence of a moduli space of supersymmetric configurations.
Only a small subspace of this moduli space is captured by the
constant configurations however; the generic configurations are
described by deformations of instanton configurations.}

If we work instead with the DBI action for the branes then the
analysis is slightly more complicated due to the nonlinearity in $F$
of
the supersymmetry condition (\ref{qthir}). But we reach a similar
conclusion, and the details are presented in Appendix \ref{AppDBI}.

\section{Tachyon condensation}
\renewcommand{\theequation}{5.\arabic{equation}}
\setcounter{equation}{0}

We now consider a D1 brane winding around the $S^1$ which we
parametrized by the coordinate $y$.  If
we take a D1-D5 system with $B=0$ then there is no force between this
`test brane' and the D1-D5 bound
state, and the test brane can sit at any distance $r$ from the bound
state without feeling any force. If we
let $B\ne 0$ then the D1-D5 system is {\it not} threshold bound, and
there is an attractive force on the
test D1 brane.\footnote{For a discussion of D-brane couplings when
$B\ne 0$ and their field theory duals see for
example
\cite{das}.}

\begin{figure}
\epsfysize=4in \epsffile{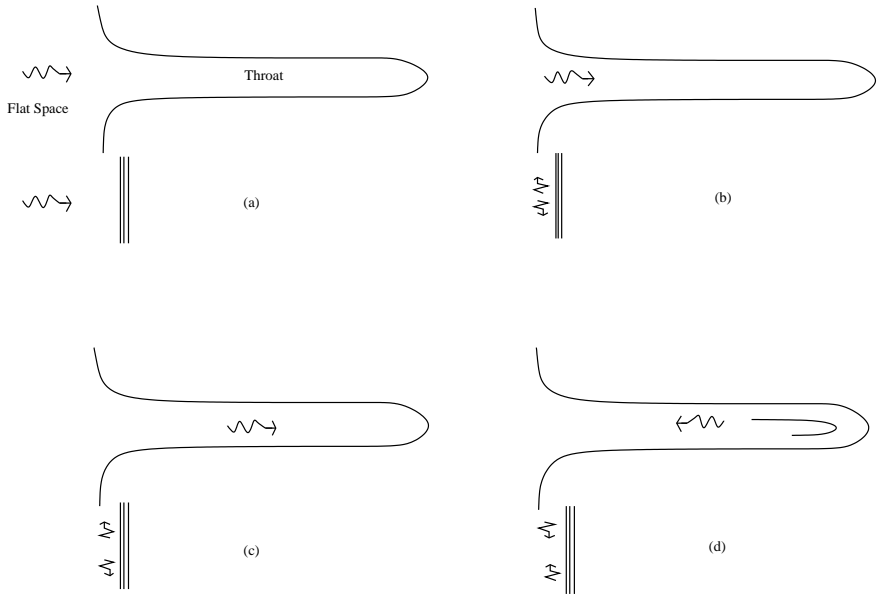}
\caption{\label{fig1}
Supergravity and field theory descriptions of the absorption of a
massless quantum by the D1-D5 bound
state. }
\end{figure}

To outline the physics we expect let us first recall the results of
\cite{lm4} where we let a
massless quantum fall into the D1-D5 `throat'.  In Fig.1 we show the
supergravity description  where
we have the metrics (\ref{fiveMetr}), and drawn below that the dual brane
description; the  branes sit in flat
spacetime.  In Fig.1(a)  the quantum is outside the throat of the
geometry, and correspondingly outside
the branes in the dual picture. In Fig.1(b) the quantum enters the
supergravity throat (with some
probability
$P_{sugra}$). In the dual picture it gets absorbed by the brane and
converted to a set of left and right
moving vibrations; the probability for this absorption process
$P_{CFT}$ is found to satisfy
$P_{CFT}=P_{sugra}$. In Fig.1(c) we find that the supergravity
quantum travels down the throat,  and in the field theory the two
excitations separate away from each
other. In Fig.1(d) the supergravity quantum {\it reflects} off the
end and returns to the start of the throat
in a time $\Delta t_{sugra}$; in the dual brane
picture the left and right moving excitations  travel around the
branes and re-collide in a time $\Delta
t_{CFT}=\Delta t_{sugra}$.

In Fig.2 we picture the corresponding process when the infalling
object is a D1 brane.
\medskip

$\bullet$\quad   In Fig.2(a) the D1
brane is far outside the throat, and in the dual picture it is
well-separated from the D1-D5
bound state. In the gravity
picture the potential energy can be found by the DBI action of the D1
brane. In the brane picture
the
potential between the D1 brane and the D1-D5 system can be found by a
1-loop computation in the open
string channel
\cite{polchinski,polchinskibook} which, at these long distances, gets
contributions
from only the lowest few modes in the
closed string channel.\footnote{One end of the open strings is at the
D1-D5 bound state, where the
boundary conditions are complicated and depend on the choice of the
Ramond ground state of the D1-D5
system.} (These potentials are just the long distance supergravity
attraction between the test brane and
the bound state.)
\medskip

$\bullet$\quad In Fig.2(b) the D1 brane is at the start of the
supergravity throat; this is the
`intermediate region' which connects flat space to the locally
$AdS_3\times S^3\times T^4$ geometry.
The DBI action of the supergravity description  yields a complicated
potential function in this region.
The force computation of \cite{polchinski}  gets contributions from
all string modes, and one must
also include multiloop processes. The D1 brane in the `brane
description' is now at $r\approx 0$.
\medskip

$\bullet$\quad In Fig.2(c) we see that in the supergravity
description the D1 brane continues deeper
into the `throat' of the geometry, where the potential energy of the
D1 brane continues to decrease.
In the dual description we expect that
lowest open string mode -- the tachyon -- begins to condense, thus
lowering the energy.
\medskip

$\bullet$\quad  In Fig.2(d) we see that the D1 brane reaches the end
of the throat, and `bounces back'.
This is the new aspect of the problem and the one that we wish to
study, and it arises from our starting
observation \cite{lm4} that the D1-D5 bound states are described by
closed throats rather than an infinite throat singular at $r=0$. The
supergravity picture  implies that in
the dual brane description the tachyon,
after going down   the potential well to some depth, climbs back up
the well (in some other direction of
field space).  We will find that the details
of the `bounce' (in the supergravity picture) depends on the choice
of Ramond ground state of the D1-D5
system, as well as the direction of infall of the D1 brane toward
the bound state.

\medskip
\begin{figure}
\epsfysize=3.5in \epsffile{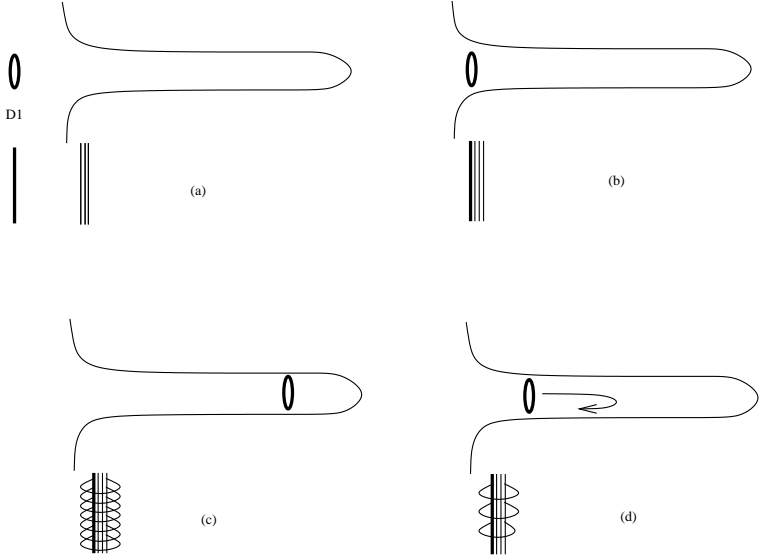}
\caption{\label{fig2}
Supergravity and field theory descriptions of the absorption of a 
D1 brane  by the D1-D5
bound state (with nonzero $B$ moduli).
}
\end{figure}

Before proceeding, we briefly compare our computation with
\cite{sw},\cite{wadia}. In \cite{sw} Seiberg
and Witten considered a D1 brane winding around the angular direction
of $AdS_3$, and computed the
potential energy needed to expand this string to different radii $r$.
In this case $B=0$, but the potential is not flat. To see the
relation with our computation, note that the
computation of \cite{sw}  was in global $AdS_3$,
which is dual to the NS sector of the D1-D5 system.  In our
computation the  D1 brane is studied in
the Ramond sector.
The two sectors are related in supergravity by a coordinate
transformation, which is mentioned  below in
(\ref{qqel}). A D1 brane that is static in the R sector is rotating on
the $S^3$ with unit velocity in the NS sector.
Such a rotating D1 brane in $AdS_3\times S^3$ is a `giant graviton'
\cite{giant} which feels no potential
against radial expansion.\footnote{Giant gravitons in $AdS_3\times
S^3$ and their interpretation in the
CFT were studied in \cite{lms}.}

In the Ramond sector we can get an attractive potential if we let $B\ne
0$, and that is the case that we are studying. Such a case was also studied in
\cite{wadia}. But in \cite{wadia} the supergravity geometry at small $r$ was an
infinite throat (the Poincare patch
with  identification $y\rightarrow y+2\pi R$), and the D1 brane fell
in towards $r=0$ without returning.

In our study we encounter a throat similar to the one in
\cite{wadia}, but we also have an end to the
throat. For a  special case of the geometry (the maximally rotating
configuration) the geometry at the end
looks (after the spectral flow coordinate transformation) just like
global $AdS_3$ times $S^3$. The fact
our D1 brane moves in the 6-dimensional space (instead of just in the
$AdS_3$) implies that for generic
initial conditions the D1 brane does not self-intersect during the `bounce'.

\subsection{The tachyon potential}

Let us assume that the profile function $ F_i(v)$ in (\ref{profile})
satisfies $\dot F_i\dot F_i=constant$. Further,
let us take $Q_1=Q_5$. Then we see from (\ref{profile}) that
\be
H(1+K)=1
\ee
and  the solution (\ref{NCSolution}) simplifies to
\bea
ds^2&=&H\left[-(dt-A_idx^i)^2+(dy+B_idx^i)^2\right]
+H^{-1}d{\vec x}d{\vec x}+dz~ dz\\
e^{2\Phi}&=&1,\quad
B_{z_1z_2}=b_{12},\quad
B_{z_3z_4}=b_{34},\nonumber\\
C^{(2)}&=&\cos\theta_1\cos\theta_2 M^{(2)},\qquad
C^{(6)}=\sin\theta_1\sin\theta_2 M^{(2)}\wedge dV,\nonumber\\
C^{(4)}&=&M^{(2)}\wedge\left[\cos\theta_1\sin\theta_2 dz_3\wedge dz_4+
\cos\theta_2\sin\theta_1 dz_1\wedge dz_2\right]
\eea
where we introduced
\be
M^{(2)}\equiv H(dt-A_idx^i)\wedge
(dy+B_idx^i)+C_{ij}dx^i dx^j.
\ee
(Note that  we have shifted $M^{(2)}$ by a constant 2-form, which amounts to
constant shifts in the RR gauge fields.  The shift of
$C^{(2)}$ must be
accompanied by a shift (\ref{C2-TF}) in $C^{(4)}$, but for the
present case $B$ is
a constant field, and the shift induced in $C^{(4)}$ is also by a constant form
which gives no field strength.)  With the restrictions we have chosen
on the solution we have
$H=0$; though this will not be the case for more general solutions we
do not expect any of the essential
physics of the infall to be different.

       To find the potential felt by the D1 brane we must dualize
$C^{(6)}$ (using the relation (\ref{ToProve}); this gives
an extra contribution to $C^{(2)}$
\be\label{ProveCDual}
{\tilde C}^{(2)}=\sin\theta_1\sin\theta_2 M^{(2)}
\ee
and the total RR 2-form becomes:
\be
{\bar C}^{(2)}=\cos(\theta_1-\theta_2)M^{(2)}
\ee

We assume that the D1-brane is in a wavefunction that is uniform on
the $T^4$, and the $T^4$
plays no further part in the analysis. The D1 brane wraps the
direction $y$, and we consider only its
center-of-mass motion -- i.e., we set to zero all vibrations of the D1 brane.
       The action is
\bea\label{DBIDef}
S=-T_1\int d^2\xi e^{-\Phi}\sqrt{-\mbox{det}(G_{ab}+B_{ab})}+
T_1\int {\bar C}^{(2)}
\eea
We choose the static gauge
\be
t=\tau,\quad y=R\sigma
\ee
where the dynamics is described by  $x^i=x^i(\tau)$.
We denote the derivatives with respect to the worldvolume variables
$\xi_0=\tau$ and $\xi_1=\sigma$ by a dot
and a prime respectively.

Then the action (\ref{DBIDef}) becomes:
\bea\label{DBI}
S=-T_1 R\int d\tau d\sigma H\sqrt{(1-A_i {\dot x}^i)^2-H^{-2}{\dot x}^2}+
T_1 R\int d\tau d\sigma H\cos(\theta_1-\theta_2)
\eea
           From this action we construct a worldsheet Hamiltonian which gives
the energy
of the configuration
\be\label{DBIEnergy}
E=2\pi T_1 R H\left[
\frac{1-A_i{\dot x}^i}{\sqrt{(1-A_i{\dot x}^i)^2-H^{-2}{\dot x}^2}}-
\cos(\theta_1-\theta_2)\right]-4\pi T_1 R\sin^2\frac{\theta_1-\theta_2}{2}
\ee
where we have added a constant so as to make the energy vanish when
the D1 brane is at rest at
spatial infinity.

To find the effective potential felt by the D1 brane we set ${\dot
x}=0$ in (\ref{DBIEnergy}), getting
\be
V(x)=-4\pi T_1 R(1-H)\sin^2\frac{\theta_1-\theta_2}{2}
\ee
Looking at the definition of $H^{-1}$ in (\ref{profile}), we note that
$0\le H < 1$ everywhere, and $H=0$ only
for points $x_i$ that are on the curve $x_i=F_i(v)$ for some $v$.  As
explained in the introduction,
we term this curve the `central curve' of
the geometry.  The minimum of the potential is
reached at all points this central curve, and we
find that the value of this minimum is universal, depending only on
the charges and $B$ field moduli but
not on the profile $F_i(v)$:
\be
V_{min}=-4\pi T_1 R\sin^2\frac{\theta_1-\theta_2}{2}
\label{qqfourt}
\ee
We will see however  that as far as the {\it dynamics} of the
tachyon is concerned, a D1 brane falling in
on a generic trajectory does not reach the minimum of the potential,
but reflects back after reaching
some other value of the potential.

The fact that the minimum (\ref{qqfourt}) does not depend on $F_i(v)$
is expected
from the fact that the binding energy
of the D1 brane to the D1-D3-D5 bound state is given only in terms of
the charges and moduli.
The potential energy (\ref{qqfourt}) is just the binding energy of the D1 brane
in the D1-D3-D5 bound state. To see this we fix and  denote the
energy of the bound state as
$M(n_1,n_{12},n_{34},n_5)$.  Then from (\ref{qqone}) we find
\bea
&&M(n_1+1,n_{12},n_{34},n_5)-M(n_1,n_{12},n_{34},n_5)\approx
\frac{dM}{dn_1}=\frac{dM}{dk_1}\nonumber\\
&&\qquad=
\left(\frac{RV}{g{\alpha'}^3}\right)^2\frac{1}{M}
\left(k_5+\frac{k_1(\alpha')^2}{V}\right)
\frac{(\alpha')^2}{V}=
\frac{R}{g{\alpha'}}\cos(\theta_1-\theta_2)
\eea
(at the last step we also used (\ref{massOne}) and (\ref{kfive})).
Then the binding energy is
$$
\delta E\equiv M(n_1+1,n_{12},n_{34},n_5)-M(n_1,n_{12},n_{34},n_5)-M(1,0,0,0)=
-\frac{2R}{g{\alpha'}}\sin^2\frac{\theta_1-\theta_2}{2}
$$
which agrees exactly with (\ref{qqfourt}) (the tension of
the D1 brane is $T_1=1/(2\pi\alpha'g)$).

We see that the test D1 brane experiences no potential in the
supergravity description
if $\theta_1=\theta_2$. From (\ref{qqtwone}) we see that this is
equivalent to ${k_{12}\over
V_{34}}={k_{34}\over V_{12}}$.  In the `brane' computation we find
that the test D1 brane experiences no
potential if
$b_{12}+2\pi\alpha'  F_{12}=b_{34}+2\pi\alpha' F_{34}$. Using
(\ref{BraneTransl}) we see that the latter
condition is the same as the condition ${k_{12}\over
V_{34}}={k_{34}\over V_{12}}$.

\subsection{Motion of an infalling D1 brane}

To explicitly find the motion of the D1 brane we take a further
subset of the above configurations --
those where $F_i(v)$ describes a circle in the $x_i$ space. These
configurations were studied in
\cite{bal}\cite{mm} and were also used in \cite{lm4} to study a
massless quantum falling down the throat. These configurations have
$\dot F_i\dot F_i=constant$, and
here
we further choose $Q_1=Q_5\equiv Q$.
The harmonic functions (\ref{profile}) are
\be\label{RingCoeff}
H^{-1}=1+\frac{Q}{f_0},\quad K=\frac{Q}{f_0},\quad
A_\phi=-\frac{Qa\sin^2\theta}{f_0},\quad
B_\psi=-\frac{Qa\cos^2\theta}{f_0},
\ee
where
\be
f_0\equiv r^2+a^2\cos^2\theta
\label{qqthir}
\ee
The flat metric on the $x_i$ space has been  written in terms of
new coordinates
\be
dx_idx_i\equiv f_0\left(\frac{dr^2}{r^2+a^2}+d\theta^2\right)+
(a^2+r^2)\sin^2\theta d\phi^2+r^2\cos^2\theta d\psi^2
\ee
and the complete metric is
\bea\label{RingMetric}
ds^2&=&\left(1+\frac{Q}{f_0}\right)^{-1}\left[
-\left(dt+\frac{Qa\sin^2\theta}{f_0}d\phi\right)^2
+\left(dy-\frac{Qa\cos^2\theta}{f_0}d\psi\right)^2
\right]\nonumber\\
&+&\left(1+\frac{Q}{f_0}\right)\left[
f_0\left(\frac{dr^2}{r^2+a^2}+d\theta^2\right)+
(a^2+r^2)\sin^2\theta d\phi^2+r^2\cos^2\theta d\psi^2
\right]+dz~dz\nonumber\\
\eea

           From the symmetry of the solution we see that we can set
${\dot\psi}={\dot\phi}=0$ for the motion of the D1 brane.  The action
(\ref{DBI}) for such
configurations becomes:
\be
S=-T_1R\int d\tau d\sigma H\left[1-\left(1+\frac{Q}{f_0}\right)^2
f_0\left(\frac{{\dot r}^2}{r^2+a^2}+{\dot\theta}^2\right)
\right]^{1/2}+
T_1 R\int d\tau d\sigma H\cos(\theta_1-\theta_2)
\label{qqfift}
\ee
We further note that
            the Lagrangian depends on
$\theta$ only through the combination $f_0=r^2+a^2\cos^2\theta$.  Thus the
derivative $\d L/\d\theta$ vanishes for $\theta=0$ and
$\theta=\pi/2$. If $\theta$ is set to either of
these values then it stays constant, and we get a 1-dimensional
problem in the variable $r$. We analyze
these two cases as they give us two physically opposite limits out of
the generic set of trajectories.
\medskip

{\bf (i) Trajectory with $\theta=0$.}
Consider the case where the total energy (\ref{DBIEnergy}) is zero. We get
\be
{\dot r}=\pm(1+\frac{Q}{r^2+a^2})^{-1}\left[1-\left(1+2\sin^2\frac{\theta_1-
\theta_2}{2}~\frac{Q}{r^2+a^2}\right)^{-2}\right]^{1/2}
\label{qqtw}
\ee
Note that ${\dot r}$   never goes to
zero ($r$ goes down to zero and then starts increasing again).  The
D1 brane travels down the throat and bounces back up,
spending only a finite time in the
throat.

To understand the details of the bounce we look at the geometry where
$a$ takes its largest value
$a=Q/R$.  The metric (reduced on $T^4$)  near the
end of the throat is  \cite{bal}\cite{mm}
\be
ds^2\approx -\frac{(r^2+a^2)dt^2}{Q}+\frac{r^2dy^2}{Q}
+
Q\left[{dr^2\over r^2+a^2}+
d\theta^2+\cos^2\theta d\psi_{NS}^2+\sin^2\theta
d\phi_{NS}^2\right]
\label{qqten}
\ee
where
\be
\psi_{NS}=\psi-\frac{y}{R}, ~~\phi_{NS}=\phi-\frac{t}{R}
\label{qqel}
\ee
We see that the metric (\ref{qqten}) describes $AdS_3\times S^3$,
with the $S^3$ described by $\theta, \psi_{NS}, \phi_{NS}$.
The spacetime $AdS_3\times S^3$ is the dual of the {\it
Neveu-Schwarz}  sector of the CFT, and we have included subscripts on
$\psi_{NS}, \phi_{NS}$ to note this fact.

The test  D1 wraps the direction $y$ and is at  constant $\psi, \phi$, but by
(\ref{qqel}) this implies that it wraps the direction $\psi_{NS}$.
The  D1 brane describing the motion (\ref{qqtw}) reaches $r=0$, but
since $\theta=0$ we see that at this point the D1 brane  wraps the
diameter of $S^3$ parametrized by $\psi_{NS}$.   Thus the D1 brane
has not shrunk to a point, and it does not self-intersect or
encounter any other singularity in the process of bouncing back to
large
$r$.

         We expect this behavior of the D1 brane to be generic in the sense that
for generic shapes of the throat end and generic initial conditions
for the D1 brane  the brane will not  shrink to a point or
self-intersect when it reaches the end of its motion down the throat.
The D1 brane is just a `giant graviton' in $AdS_3\times S^3$, and it
is
important to note that its motion is reliably given by the DBI action
plus Chern-Simons term when $n_1, n_5>>1$ (with other
parameters held fixed).

\bigskip

{\bf (ii) Trajectory with $\theta=\pi/2$.}
            In this case we find:
\be
{\dot r}=\pm\frac{\sqrt{r^2+a^2}}{r}
(1+\frac{Q}{r^2})^{-1}\left[1-\left(1+2\sin^2\frac{\theta_1-
\theta_2}{2}~\frac{Q}{r^2}\right)^{-2}\right]^{1/2}
\ee
As $r$ goes to zero we find ${\dot r}\sim -ra/Q$, so the time to
reach $r=0$ diverges.

The D1 brane again wraps the direction parametrized by $\psi_{NS}$,
but since now $\theta=\pi/2$ we find that the brane  shrinks
to a point as $r\rightarrow 0$.  Note that since $\phi$ is constant
we have $d\phi_{NS}/dt=1$.  Thus  as
$t\rightarrow\infty$ the D1 brane settles down to a pointlike quantum
traveling at the speed of light along the diameter of the $S^3$
parametrized by $\phi_{NS}$. The  D1 brane becomes, at late
times, an {\it ordinary} graviton rather than a giant graviton.

We expect that this diverging
time will be found for an exceptional set of initial conditions.
        From (\ref{qqthir}) we see that the curve $r=0, ~\theta=\pi/2$
is the `central curve' for the geometry under consideration, i.e. the
curve occupied by the profile $F_i$ in (\ref{profile}). (Recall that points
on
this curve were the centers  of a Kaluza-Klein  geometry in
directions transverse to the curve \cite{lmm}.)
We see that if the initial conditions on the D1 brane are such that
they send the D1 brane  straight into a point on the `central curve'
then
the D1 brane settles down to a pointlike graviton as
$t\rightarrow\infty$ instead of bouncing back. Correspondingly, in
the dual brane
picture the tachyon settles down to the minimum value
(\ref{qqfourt}) of the potential. (The potential takes this value
(\ref{qqfourt})
at all points of the central curve but nowhere else.)

\bigskip

{\bf (iii) Trajectory for $a=0$.}
As discussed in \cite{lm4} the parameter $a$ can go down to very
small values; it is prevented from
vanishing only by the fact that the minimum angular momentum is
$j=\hbar/2$ for the quantum state
dual to the supergravity solution. To study D1 brane infall for these
geometries with small $a$ we set
$a=0$.  Now we can set $\theta=\theta_0, \phi=\phi_0, \psi=\psi_0$
and the radial motion is described by
\be
{\dot r}=\pm
(1+\frac{Q}{r^2})^{-1}\left[1-\left(1+2\sin^2\frac{\theta_1-
\theta_2}{2}~\frac{Q}{r^2}\right)^{-2}\right]^{1/2}
\ee
We note that in this case the travel time from $r\sim\sqrt{Q}$ to
$r\sim a$ (where the throat ends)
is $\sim 1/a$. Thus in D1-D5 bound states with small $a$  the tachyon
bounces back after a minimum
time $\sim 1/a$.

\bigskip

Thus in the gravity picture the reason for the `bounce' of the D1
brane is quite simple: when the D1 brane
reaches the end of the throat it generically avoids falling onto the
`central curve'. This is somewhat
similar to having a nonzero impact parameter in the process of infall
towards a point singularity at $r=0$;
in this latter case the infalling object would also return to larger
values of $r$.\footnote{In a recent paper
\cite{branonium} the motion of a D brane near a
cluster of anti-branes was  considered, and it was
noted that angular momentum would prevent
quick annihilation.} What is interesting in  our
problem is that the D1-D5 bound state has an
inherent nontrivial structure, and it becomes a
complicated question to determine
      which trajectories of the D1 brane correspond to `zero impact
parameter'.  The choice of D1-D5 bound state
determines the central curve of the geometry, and the direction of
infall of the D1 brane (i.e. the choice of
$\theta_0,
\psi_0,
\phi_0$) also determines the details of the bounce.\footnote{In the
field theory
picture it is na\"{\i}ve to think of the transverse displacements of
the D1 brane as
parametrized by points in ${\cal R}^4$.  For the D0-D4 system it was shown in
\cite{dkps} that 1-loop effects in open string theory make the
moduli space corresponding to these displacements singular at $r=0$.
This effect is
related to the fact that in the supergravity picture even when the
test D1 brane falls
deep down the throat we still have to specify $\theta,\phi,\psi$ to
specify its state.}

   It would be interesting to study the above
phenomena directly in the dual field
theory.\footnote{Some steps
towards identifying the tachyon when $B\ne 0$
were taken in
\cite{wadia}; the D1-D5 system with $B$ field was
also studied in \cite{mik}.}

\section{Discussion}
\renewcommand{\theequation}{6.\arabic{equation}}
\setcounter{equation}{0}

We have constructed metrics dual to bound states of  D1-D3-D5 branes
(with nonzero $B$),
and analyzed the
motion of a test  D1 brane in these geometries. We found that the D1
brane generically
bounces back from the end of the `throat'. In the dual field theory
description this implies
that the tachyon starts by condensing towards the bottom of its
potential, but can then
bounce back up the potential hill. We now comment on the physics of
this process.

The circle $S^1$ parametrized by the coordinate $y$ has a length $2\pi R$ (at
$r\rightarrow\infty$). The dual  description is given by a
1+1 dimensional
field theory with the spatial direction  a  circle of length $2\pi
R$. If we let
$R\rightarrow\infty$ with all other parameters like $g,\alpha', n_1,
n_5, b_{ij}$ fixed,
then in the supergravity computation we find that the `bounce back' time
$t_{bounce}$ for the test D1 brane goes to infinity.  In the dual
brane description we
conclude that if the CFT is {\it not} compactified to a circle (i.e.
if the CFT lives on ${\cal R}$,
the real line) then the tachyon settles down towards its minimum, and
does not bounce
back in any finite time.

But for applications to black holes we need to consider the D1-D5
system compactified
on   a circle, and in this case it was shown in \cite{lm4} that the
throat  of the geometry
ends after a finite distance, and infalling quanta are reflected back
from this end. The
length $R$ scales out from the final quantities of interest; for
example it was shown in
\cite{lm5} that the `horizon area' obtained by coarse graining over
the different possible
endings of the throat gave the Bekenstein entropy of the 2-charge
system. To consider
questions like the fate of a D1 brane falling into a black hole, we
must look at tachyon
condensation when $R$ is {\it finite}.

The low energy field theory of the D1-D5  bound state   can be
written as   a 1+1
dimensional sigma model \cite{stromingervafa,dijkgraaf,sw}, with target space
a deformation of the  orbifold
$(T^4)^N/S_N$ (the symmetric product of
$N=n_1n_5$ copies of $T^4$). The action of twist operators
$\sigma_{n_i}$ leads to the
ground  states being characterized by `component strings' of
different lengths $2\pi R
n_i$, and each component string also carries a $SU(2)\times SU(2)$
spin under the
rotation group of the $S^3$ surrounding the branes
\cite{lm1,lm4}.
The orders of the twists $\{ n_i \}$ and the spins determine the
shape of the end of the
throat.

   The supergravity
computation tells
us that after the  effects of these twists,
spins etc. are taken
into account, the tachyon generically bounces back after reaching
some point close to its
minimum energy point.\footnote{Note that when we take $R\rightarrow \infty$ the
magnitudes of the spins etc. stay the same, so in the field theory
the effects of these
spins, twists etc.   gets `diluted'. As a result we find
that the  bounce back time of the
tachyon diverges.} The supergravity computation is reliable once we let
$n_1, n_5$ be large (for other parameters fixed); for example in the case
$a=Q/R$ in
the metrics (\ref{NCSolution}) with coefficients (\ref{RingCoeff})
we have seen that the D1 brane at the end of the
throat is just a
giant graviton in $AdS_3\times S^3$, and we know that the
backreaction of the giant
graviton is small for $n_1, n_5\gg 1$.

It is interesting that there are specific choices of initial
conditions, at least for the metrics parameterized by harmonic functions
(\ref{RingCoeff}), where the D1 brane asymptotically settles down to a
pointlike object -- a
`graviton', as opposed to a giant graviton. This happens if the D1
brane evolves such as to
shrink down to a point on the `central curve' of the geometry.

Note that the initial state of the test D1 brane is chosen to be
translationally invariant
in $y$, and the  geometries  (\ref{NCSolution}) are also
translationally invariant in $y$.
Thus the classical motion of the D1 brane in any of these geometries
will not generate
vibrations of the test D1 brane, and the dynamical problem is limited
to the center of
mass motion of this brane.    For generic shapes of the central curve
and generic
direction of infall
   the D1 brane need not bounce back to the start of the throat after reflecting
off the end; it may stay trapped for long times near
the end of the  throat as it  moves
in the transverse 4-dimensional space $x_i, i=1\dots 4$.\footnote{The
D1 brane can
radiate energy by higher-loop processes as it moves, and so will
ultimately settle to
the bottom of the potential (a similar observation was made in
\cite{branonium}).}
This would be similar to the situation discussed in
\cite{lm5} for the evolution of a massless quantum
-- it was argued that the quantum stays trapped
near the end of the throat for long times when
the central curve has a complicated shape.

\section*{Acknowledgements}

This work was supported in part by NSF grant
PHY--0070928 and by DOE grant DE-FG02-91ER-40690.
We thank Sumit Das and T.A. Tran for helpful discussions.

\appendix

\section{Derivation of the Solution}
\renewcommand{\theequation}{A.\arabic{equation}}
\setcounter{equation}{0}

\subsection{Acting on directions (z$_1$,z$_2$)}

We start with the solution (\ref{fiveMetr})--(\ref{five}), and perform a
T-duality along
$z_1$, obtaining
\bea
ds^2&=&\sqrt{\frac{H}{1+K}}\left[-(dt-A_idx^i)^2+(dy+B_idx^i)^2\right]
               +\sqrt{\frac{1+K}{H}}
d{\vec x}d{\vec x}\nonumber\\
&+&  \frac{1}{\sqrt{H(1+K)}} dz_1^2+\sqrt{H(1+K)}
\left[dz_2^2+(dz_3^2+dz_4^2)\right]\\
e^{2\Phi'}&=&\sqrt{H(1+K)}
             \eea
             \bea
C'^{(3)}_{\m\a z_1}&=&C^{(2)}_{\m\a} \label{tz1c3}
             \eea
\\
\noindent
We now perform the rotation
             \bea
             z_1&=&\cos\th_1 z'_1-\sin\th_1 z'_2 \nn\\
             z_2&=&\sin\th_1 z'_1+\cos\th_1 z'_2
             \eea
which gives
             \bea
ds^2&=&\sqrt{\frac{H}{1+K}}\left[-(dt-A_idx^i)^2+(dy+B_idx^i)^2\right]
               +\sqrt{\frac{1+K}{H}}
d{\vec x}d{\vec x}\nonumber\\
&+& \left[
\frac{\cos^2\th_1}{\sqrt{H(1+K)}}+\sqrt{H(1+K)}\sin^2\th_1
\right]{dz'}_1^2   \nn\\
&+&\left[\frac{\sin^2\th_1}{\sqrt{H(1+K)}}
             + \sqrt{H(1+K)}\cos^2\th_1\right]{dz'}_2^2   \nn\\
&-&2\sin\th_1 \cos\th_1\left[\frac{1}{\sqrt{H(1+K)}}
             - \sqrt{H(1+K)}\right]{dz'}_1{dz'}_2   \nn\\
             &+&\sqrt{H(1+K)}\left[(dz_3^2+dz_4^2)\right]
             \eea
             \bea
             e^{2\Phi'}&=&\sqrt{H(1+K)},\quad
             \eea
             \bea
C'^{(3)}_{\m\a z'_1}&=&\cos\th_1 C'^{(3)}_{\m\a z_1}
                                 +\sin\th_1 C'^{(3)}_{\m\a z_2}
                                =\cos\th_1 C'^{(3)}_{\m\a z_1}
                                =\cos\th_1 C^{(2)}_{\m\a} \nn\\
C'^{(3)}_{\m\a z'_2}&=&-\sin\th_1 C'^{(3)}_{\m\a z_1}
                                 +\cos\th_1 C'^{(3)}_{\m\a z_2}
                                =-\sin\th_1 C'^{(3)}_{\m\a z_1}
                                =-\sin\th_1 C^{(2)}_{\m\a}
                              \label{rz1z2c3}
             \eea
We now perform another T-duality along $z_1'$. Note that
\bea
             C''^{(2)}_{\a\b}=C'^{(3)}_{\a\b z'_1}=\cos\th_1C'^{(3)}_{\m\a z_1}
                             =\cos\th_1 C^{(2)}_{\a\b}
             \eea
             \bea
C''^{(4)}_{\m\n z'_2 z'_1}&=&C'^{(3)}_{\m\n
z'_2}-3\frac{C'^{(3)}_{[\m\n |z'_1}G_{z'_2]z'_1}}{G'_{z'_1 z'_1}}
=C'^{(3)}_{\m\n z'_2}-\frac{C'^{(3)}_{\m\n
z'_1}G_{z'_2z'_1}}{G'_{z'_1 z'_1}} \nn\\
&=& -\sin\th_1 C^{(2)}_{\m\n}\,\frac{H(1+K)}{h_1}
             \eea
and we get the solution (\ref{NCSolution}) with $\theta_2=0$.

\subsection{(z$_3$,z$_4$)-directions}

We drop primes on the variables obtained above, and  perform a
similar set of operations on
$z_3, z_4$.
T$_{z_3}$ gives
             \bea
ds^2&=&\sqrt{\frac{H}{1+K}}\left[-(dt-A_idx^i)^2+(dy+B_idx^i)^2\right]
+\sqrt{\frac{1+K}{H}}
d{\vec x}d{\vec x}\nonumber\\
&+&\frac{\sqrt{H(1+K)}}{h_1}(dz_1^2+dz_2^2)
+\frac{1}{\sqrt{H(1+K)}}dz_3^2+\sqrt{H(1+K)}dz_4^2
             \eea
             \bea
             e^{2\Phi'}&=&\frac{\sqrt{H(1+K)}}{h_1},\quad
B'_{z_1z_2}=B_{z_1z_2}=\frac{\sin\theta_1\cos\theta_1}{h_1}(1-(1+K)H)
             \eea
             \bea
C'^{(3)}_{\m\a z_3}&=&C^{(2)}_{\m\a},\quad C'^{(5)}_{\m\n\rho\a
z_3}=C^{(4)}_{\m\n\rho\a} \label{tz3c3c5}
             \eea
The rotation
             \bea
             z_3&=&\cos\th_2 z'_3-\sin\th_2 z'_4 \nn\\
             z_4&=&\sin\th_2 z'_3+\cos\th_2 z'_4
             \eea
gives
             \bea
ds^2&=&\sqrt{\frac{H}{1+K}}\left[-(dt-A_idx^i)^2+(dy+B_idx^i)^2\right]
               +\sqrt{\frac{1+K}{H}}
d{\vec x}d{\vec x}\nonumber\\
&+&\frac{\sqrt{H(1+K)}}{h_1}\left[(dz_1^2+dz_2^2)\right]\nn\\
             &+& \left[
             \frac{\cos^2\th_2}{\sqrt{H(1+K)}}+\sqrt{H(1+K)}\sin^2\th_2
\right]{dz'}_3^2   \nn\\
&+&\left[\frac{\sin^2\th_2}{\sqrt{H(1+K)}}
             + \sqrt{H(1+K)}\cos^2\th_2\right]{dz'}_4^2   \nn\\
&-&2\sin\th_2 \cos\th_2\left[\frac{1}{\sqrt{H(1+K)}}
             - \sqrt{H(1+K)}\right]{dz'}_3{dz'}_4
             \eea
             \bea
             e^{2\Phi'}&=&\frac{\sqrt{H(1+K)}}{h_1},\quad
B'_{z_1z_2}=\frac{\sin\theta_1\cos\theta_1}{h_1}(1-(1+K)H)
             \eea
             \bea \label{rz3z4c3c5}
C'^{(3)}_{\m\a z'_3}&=&\cos\th_2 C'^{(3)}_{\m\a z_3}
                                 +\sin\th_2 C'^{(3)}_{\m\a z_4}
                                =\cos\th_2 C'^{(3)}_{\m\a z_3}
                                =\cos\th_2 C^{(2)}_{\m\a} \nn\\
C'^{(3)}_{\m\a z'_4}&=&-\sin\th_2 C'^{(3)}_{\m\a z_3}
                                 +\cos\th_2 C'^{(3)}_{\m\a z_4}
                                =-\sin\th_2 C'^{(3)}_{\m\a z_3}
                                =-\sin\th_2 C^{(2)}_{\m\a} \nn\\
C'^{(5)}_{\m\n\rho\a z'_3}&=&\cos\th_2 C'^{(5)}_{\m\n\rho\a z_3}
                                 +\sin\th_2 C'^{(5)}_{\m\n\rho\a z_4}
                                =\cos\th_2 C'^{(5)}_{\m\n\rho\a z_3}
                                =\cos\th_2 C^{(4)}_{\m\n\rho\a} \nn\\
C'^{(5)}_{\m\n\rho\a z'_4}&=&\!\!-\sin\th_2 C'^{(5)}_{\m\n\rho\a z_3}
                                 +\cos\th_2 C'^{(5)}_{\m\n\rho\a z_4}
                                =\!\!-\sin\th_2 C'^{(5)}_{\m\n\rho\a z_3}
                                =\!\!-\sin\th_2 C^{(4)}_{\m\n\rho\a}
             \eea
Finally,  T$_{z'_3}$ gives the solution (\ref{NCSolution}).

\section{T-duality formulae}
\renewcommand{\theequation}{B.\arabic{equation}}
\setcounter{equation}{0}

In this paper we perform T dualities following the notation of \cite{johnson}.
Let us summarize the relevant formulae.
We call the T-duality direction $s$. For NS--NS fields, one has
\bea
G'_{ss}=\frac{1}{G_{ss}},\qquad
e^{2\Phi'}=\frac{e^{2\Phi}}{G_{ss}},&&\ G'_{\mu s}=\frac{B_{\mu
s}}{G_{ss}},\qquad B'_{\mu s}=\frac{G_{\mu s}}{G_{ss}}
\nonumber\\
G'_{\mu \nu}=G_{\mu \nu}-\frac{G_{\mu s}G_{\nu s}-B_{\mu s}B_{\nu s}}{G_{ss}},
&&\
B'_{\mu \nu}=B_{\mu \nu}-\frac{B_{\mu s}G_{\nu s}-G_{\mu s}B_{\nu s}}{G_{ss}},
\eea
while for the RR potentials we have:
\bea\label{RRTDual1}
{C'}^{(n)}_{\mu\dots\nu\alpha s}&=&C^{(n-1)}_{\mu\dots\nu\alpha}-
(n-1)\frac{C^{(n-1)}_{[\mu\dots\nu|s}G_{|\alpha]s}}{G_{ss}},\\
{C'}^{(n)}_{\mu\dots\nu\alpha\beta}&=&C^{(n+1)}_{\mu\dots\nu\alpha\beta s}
+nC^{(n-1)}_{[\mu\dots\nu\alpha}G_{\beta]s}
+n(n-1)\frac{C^{(n-1)}_{[\mu\dots\nu|s}B_{|\alpha |s}G_{|\beta ]s}}{G_{ss}}.
\eea

\section{Mass Formulae for D0-D2-D4 system}
\label{AppMass}
\renewcommand{\theequation}{C.\arabic{equation}}
\setcounter{equation}{0}

We consider Type IIA string theory, and regard it as a dimensional
reduction of 11-dimensional  M-theory.
We compactify the IIA theory on a torus, and wrap the D-branes on
directions along this torus. For a supersymmetric bound
state of  D0, D2 and D4 branes we write
\cite{towns,pioline}
\begin{equation}
\gm\epsilon = M \epsilon \label{appendixzero}
\end{equation}
where
\begin{equation}
\gm \equiv Z \gm_{0s}+ \frac{1}{2!}Z^{ij}
\gm_{0ij}+\frac{1}{4!}Z^{ijkl}\gm_{0ijkls}
\end{equation}

The index $s$ represents the compact $11-$direction of M-theory,
while the other indices take values
along the compact torus.
             Expanding out the equation above we find
\dm
\gm^{2}= Z^{2} \gm_{0s}\gm_{0s} + \frac{1}{2}Z Z^{ij}
\ac{\gm_{0s}}{\gm_{0ij}} + \frac{1}{4!} Z Z^{ijkl}
\ac{\gm_{0s}}{\gm_{0ijkls}} +
\edm
\be
             \frac{1}{8} Z^{ij}Z^{kl}\ac{\gm_{0ij}}{\gm_{0kl}} + \frac{1}{2.4!}
Z^{ij}Z^{klmn}\ac{\gm_{0ij}}{\gm_{0klmns}} + \frac{1}{2(4!)^{2}}
Z^{ijkl}Z^{mnpq}\ac{\gm_{0ijkls}}{\gm_{0mnpqs}}
\ee
The first five terms simplify to
\be
Z^{2}+\frac{1}{12} Z Z^{ijkl} \gm_{ijkl} +\frac{|Z^{ij}|^{2}}{2} -
\frac{ Z^{\left[ ij\right.}Z^{\left. kl\right]} \gm_{ijkl}}{12} -
\frac{1}{3}Z^{i[j}Z^{npq]i}\gm_{jnpqs} \label{appendixone}
\ee
The last term can be written as
\be
\frac{1}{2} Z^{ijkl}Z^{mnpq}\ac{\gm_{ijkl}}{\gm_{mnpq}}_{i<j<k<l,\ m<n<p<q}
\ee
which simplifies in general to
\be
\frac{1}{4!} |Z^{ijkl}|^{2}
-\frac{1}{4}Z^{ij[kl}Z^{pq]ij}\gm_{klpq}+\frac{1}{(4!)^{2}}
Z^{[ijkl}Z^{mnpq]}\gm_{ijklmnpq}
\label{qone}
\ee

In our D1-D5 system we have IIB theory compactified on $T^5=T^4\times
S^1$, and thus we consider branes in the the T-dual
IIA theory  wrapped on at most 5 compact directions.  With the
indices $i,j\dots$ limited to 5 possible values, we find that  the
last two terms in (\ref{qone})  are identically zero, and we get
\be
\gm^{2}\epsilon= \left(Z^{2}+  \frac{1}{12} Z Z^{ijkl} \gm_{ijkl}
+\frac{|Z^{ij}|^{2}}{2} - \frac{ Z^{\left[ ij\right.}Z^{\left.
kl\right]} \gm_{ijkl}}{12} - \frac{1}{3} Z^{i[j}Z^{npq]i}\gm_{jnpqs}
+\frac{1}{4!} |Z^{ijkl}|^{2} \right)\epsilon \label{appendixthree}
\ee

Defining
\be
k^{ijkl} \equiv 2(Z Z^{ijkl}- Z^{\left[ ij\right.}Z^{\left.
kl\right]}),\ \ k'^{ijkl}\equiv Z^{i[j}Z^{npq]i}
\ee
we find that 1/2 BPS  configurations are obtained for
\be
k^{ijkl}=k'^{ijkl}=0 \label{appendixtwo}
\ee
and the mass for such configurations is given by
\be
\gm^{2}\epsilon \equiv M_{0}^{2}\epsilon= \left(Z^{2}+\frac{1 }{2}
|Z^{ij}|^{2}  +\frac{1}{4!} |Z^{ijkl}|^{2}\right)\epsilon\ee

In the present paper we have wrapped branes only on $T^4$ out of the
$T^4\times S^1$, and thus the indices $i,j\dots$ are
limited to only 4 possible values. Then $ k'^{ijkl}=0$, and eqn.
(\ref{appendixthree}) can be rewritten as
\be
(M^2-M_{0}^2)^2\epsilon=
\left(\frac{k^{ijkl}\gm_{ijkl}}{4!}\right)^{2}\epsilon =
\frac{1}{4!}|k^{ijkl}|^2
\ee
which yields
\be
M^2=  \left(Z^{2}+\frac{1 }{2} |Z^{ij}|^{2}  +\frac{1}{4!}
|Z^{ijkl}|^{2}\right)
+ \sqrt{\frac{|k^{ijkl}|^2}{4!}}
\ee
With $k^{ijkl}\ne 0$ this is the mass formula for 1/4 BPS states.

We now relate the $Z$ variables to the number of branes in the state.
Let the $T^4$ be rectangular with sides $L_i, i=1\dots 4$.
Define the dimensionless parameters
\be
\ell_{i} = \frac{L_{i}}{2\pi\sqrt{\alpha'}}, ~~~
v= \ell_{1}\ell_{2}\ell_{3}\ell_{4}
\ee
By writing the mass expected when each kind of brane is present by
itself, we get the identifications
\be
Z= \frac{1}{g\sqrt{\alpha' }}n_{0},\ Z^{ij}=
\frac{\ell_{i}\ell_{j}}{g\sqrt{\alpha'}} n_{ij},\ Z^{ijkl}=
\frac{\ell_{i}\ell_{j}\ell_{k}\ell_{l}}{g\sqrt{\alpha'}} n_{ijkl}
\ee

We are interested in the particular case of D0-D2-D4 system where
we have  $D4$ branes wrapped along the $1234$
directions of $T^4$,  $D2$ branes wrapped  along the $12$  and $34$
directions, as well as some
$D0$ branes. Let
$n_0$ be the number of zero branes, $n_{12}, n_{34}$ the number of
two branes in direction $12$ and $34$ respectively  and
$n_{4}$ the number of four branes. Then
\be
M^{2}=
\frac{1}{g^2\alpha'}\left((\ell_{1}\ell_{2}n_{12}\mp\ell_{3}\ell_{4}n_
{34})^{2}+ (n_{0}\pm\ell_{1}\ell_{2}\ell_{3}\ell_{4}n_{4})^{2}\right)
\label{qtwo}
\ee
where we choose the upper signs if $n_0n_4-n_{12}n_{34}>0$ and the
lower signs if
$n_0n_4-n_{12}n_{34}<0$.

The above relations are for vanishing value of the $B$ field. For
$B\ne 0$ we can obtain the mass by rewriting the charges of the
bound state in terms of  the (matrix-valued) field strength $F$ on
the D4 branes, and then noting that all quantities depend only on
the combination   $F+ \frac{B}{2\pi\alpha'}$.  We have
\begin{eqnarray}
             n_{0}&=&\frac{L_{1}L_{2}L_{3}L_{4}}{(2\pi)^{2}}Tr(F_{12}F_{34})\\
             n_{12}&=& \frac{L_{3}L_{4}}{2\pi}Tr(F_{34}) \\
n_{34}&=& \frac{L_{1}L_{2}}{2\pi}Tr(F_{12}) 
\end{eqnarray}
Thus for $B\ne 0$ we must make the replacements
\bea
n_{12} &\rightarrow& n_{12} + n_{4}\ell_{3}\ell_{4}b_{34}\nonumber\\
n_{34} &\rightarrow& n_{34} + n_{4} \ell_{1}\ell_{2}b_{12}\nonumber\\
n_{0} &\rightarrow& n_{0}+ \ell_{1}\ell_{2}b_{12}n_{12} +
\ell_{3}\ell_{4}b_{34}n_{34} +  n_{4} \ell_{1}\ell_{2}
\ell_{3}\ell_{4} b_{12}b_{34}
\eea
Substituting these in (\ref{qtwo}) (and regrouping terms) we get
\dm
M^{2}=
\frac{1}{g^2\alpha'}\left(\left(\ell_{1}\ell_{2}n_{12}\mp
\ell_{3}\ell_
{4}n_{34}+ n_{4}v (b_{34}\mp b_{12})\right)^{2}+\right.
\edm
\be
\left.\left(n_{0}+ v(b_{12}b_{34}\pm 1)
n_{4}+\ell_{1}\ell_{2}b_{12}n_{12} +
\ell_{3}\ell_{4}b_{34}n_{34}\right)^{2}\right)
\ee

\section{Supersymmetric brane configurations using the DBI action}
\label{AppDBI}
\renewcommand{\theequation}{D.\arabic{equation}}
\setcounter{equation}{0}

We assume that $n_4$ is even, and look for a specific supersymmetric
configuration to establish that the bound state can reach the
supersymmetric mass bound. Let the field strength have values
$F_{12}, F_{34}$ on half of the D4 branes, and
values
$F_{12}', F_{34}'$ on the other half.  Then the constraints are
\be
({n_4\over 2})(F_{12}+F_{12}')=({2\pi\over L_1L_2}) n_{34}
\label{wone}
\ee
\be
({n_4\over 2})(F_{34}+F_{34}')=({2\pi\over L_3L_4}) n_{12}
\label{wtwo}
\ee
\be
({n_4\over 2})(F_{12}F_{34}+F_{12}'F_{34}')=({(2\pi)^2\over
L_1L_2L_3L_4}) n_{0}
\label{wthree}
\ee
The supersymmetry condition (\ref{qthir}) is
\be
{F_{12}\pm F_{34}\over 1\mp (2\pi\alpha')^2 F_{12}F_{34}
}={F'_{12}\pm F'_{34}\over 1\mp (2\pi\alpha')^2 F'_{12}F'_{34} }
\label{wfour}
\ee

Define
\be
\alpha=\pm ({2\over n_4})^2{1\over v}~[n_{12}n_{34}-n_0n_4 ]
\label{wsix}
\ee
\be
\beta= ({2\over n_4})^2{1\over
v}~[\pm(n_{12}n_{34}-n_0n_4)+({l_3l_4\over l_1l_2})n_{34}^2+v n_4^2]
\label{wseven}
\ee
\be
\gamma=({2\over n_4})^2{1\over
v}~[\pm(n_{12}n_{34}-n_0n_4)+({l_1l_2\over l_3l_4})n_{12}^2+v n_4^2]
\label{weight}
\ee

Then the solution to (\ref{wone}), (\ref{wtwo}), (\ref{wthree}),
(\ref{wfour}) is
\be
(2\pi\alpha')F_{12}=\pm{1\over n_4}~\left[({1\over l_1l_2})n_{34}+{1\over
\sqrt{v}}~\sqrt{\alpha\beta\over \gamma}\right]
\ee
\be
(2\pi\alpha')F'_{12}=\pm{1\over n_4}~\left[({1\over l_1l_2})n_{34}-{1\over
\sqrt{v}}~\sqrt{\alpha\beta\over \gamma}\right]
\ee
\be
(2\pi\alpha')F_{34}=\pm{1\over n_4}~\left[({1\over l_3l_4})n_{12}-{1\over
\sqrt{v}}~\sqrt{\alpha\gamma\over \beta}\right]
\ee
\be
(2\pi\alpha')F'_{34}=\pm{1\over n_4}~\left[({1\over l_3l_4})n_{12}+{1\over
\sqrt{v}}~\sqrt{\alpha\gamma\over \beta}\right]
\ee

We note from (\ref{wsix}),  (\ref{wseven}), (\ref{weight}) that if
$(n_{12}n_{34}-n_0n_4)\ge 0$ then we can use the upper sign in
these relations, and obtain a real solution for $F$. If
$(n_{12}n_{34}-n_0n_4)\le 0$ then we use the lower sign and again get
a real
solution
$F$. Thus we have a supersymmetric configuration for all values of the charges.

\end{document}